\documentclass{article}

\usepackage{arxiv}

\usepackage[utf8]{inputenc} 
\usepackage[T1]{fontenc}    
\usepackage{hyperref}       
\usepackage{url}            
\usepackage{booktabs}       
\usepackage{amsfonts}       
\usepackage{nicefrac}       
\usepackage{microtype}      
\usepackage{lipsum}		
\usepackage{graphicx}

\title{The PHOTON Wizard - Towards Educational Machine Learning Code Generators}

\date{}

\author{
  Ramona Leenings\textsuperscript{*,1,2}\thanks{\newline*These authors contributed equally. \newline 1 - Department of Mental Health, University of M\"{u}nster, Germany. \newline 2 - Faculty of Mathematics and Computer Science, University of M\"{u}nster, Germany.} \\
  \And Nils Ralf Winter\textsuperscript{*,1}
  \AND Kelvin Sarink\textsuperscript{1}
  \And Jan Ernsting\textsuperscript{1}
  \And Xiaoyi Jiang\textsuperscript{2}
  \And Udo Dannlowski\textsuperscript{1}
  \And Tim Hahn\textsuperscript{1}
}

\begin{document}
\maketitle

\begin{abstract}
Despite the tremendous efforts to democratize machine learning, especially in applied-science, the application is still often hampered by the lack of coding skills. As we consider programmatic understanding key to building effective and efficient machine learning solutions, we argue for a novel educational approach that builds upon the accessibility and acceptance of graphical user interfaces to convey programming skills to an applied-science target group. We outline a proof-of-concept, open-source web application, the PHOTON Wizard, which dynamically translates GUI interactions into valid source code for the Python machine learning framework PHOTON. Thereby, users possessing theoretical machine learning knowledge gain key insights into the model development workflow as well as an intuitive understanding of custom implementations. Specifically, the PHOTON Wizard integrates the concept of Educational Machine Learning Code Generators to teach users how to write code for designing, training, optimizing and evaluating custom machine learning pipelines. 
\end{abstract}


\section{Introduction}
\label{introduction}

The field of Machine Learning (ML) has drawn unparalleled interest from science and industry alike. Recent efforts to democratize the accessibility \cite{GoogleAi,wharton,MedicineAi,MicrosoftAi,OpenML} yielded both free educational content and an increasing number of well-maintained, open-source ML software libraries \cite{scikitlearn, scikitlearnapi,tensorflow,keras,imblearn,skopt,smac}. However, their effective use still requires substantial programming skills, rendering many breakthroughs inaccessible to researchers in the applied sciences.

In general, educational resources for acquiring the requested skills already exist \cite{coursera,deeplearningai,udemy,mmmastery}. A plethora of online courses introducing different programming languages and basic programming concepts are freely available. Likewise, many online tutorials showcase function and syntax of ML software libraries and explain their usage. For real-life ML projects, however, the exemplary solutions are usually insufficient: In order to handle common obstacles such as class imbalance, small sample sizes, and the integration of different data modalities, suitable algorithms for the problem at hand must not only be identified, but integrated across diverse software libraries and coding paradigms. In addition, the basic ML workflow - including training, evaluation and hyperparameter optimization - needs to adhere to best-practice guidelines potentially requiring e.g. nested cross-validation and hyperparameter optimization. Finally, the data format must be adapted to suit different library-specific requirements. Thus, for an inexperienced user the task of implementing efficient and effective ML solutions suitable for the learning task at hand may appear insurmountable.

While providing an easy-to-use alternative, graphical user interfaces (GUIs) \cite{gui} come at a cost: First, in comparison to the vast number of options offered in common Python libraries, the functionality accessible via the GUI is usually very limited. Especially in an applied-science context, covering all suitable elements such as learning algorithms, modality-specific processing, performance metrics, hyperparameter optimization algorithms and cross-validation strategies in one particular application appears impossible. Second, as GUI-based ML tools are self-contained, users depend on the availability and maintenance of the software system. Furthermore, the implementation of methodological advances is at the sole discretion of the product developer, hampering innovation especially in a scientific context.

In contrast, we envision a system that utilizes the advantages of a GUI to convey the necessary coding skills. We propose an \textit{Educational Machine Learning Code Generator} framework that mirrors existing GUI-based systems and enables instant access to ML, but in addition translates each GUI interaction into valid source code. Instant feedback lets users experience the impact of their choices on the underlying code, intuitively conveying insights into the emergence of custom solutions. While eleviating the burden of coding and structuring the ML workflow, a basic understanding of theoretical ML concepts - e.g. the choice of algorithms or cross-validation - is however required. The concept of \textit{Educational Machine Learning Code Generators} aims at minimizing the entry barrier for users possessing theoretical knowledge about fundamental ML concepts, in order to enable them to conduct analysis right away. We thereby hope to increase user motivation and to mitigate the conflict of accessibility versus flexibility.

While generally possible with any ML library or toolbox, we choose to build our Educational Machine Learning Generator based on the Python library PHOTON \cite{photon}. PHOTON introduces a layer of abstraction to the ML model development workflow, enabling the design of advanced machine learning pipelines and encapsulating several existing machine learning toolboxes \cite{scikitlearn,tensorflow,keras,imblearn,skopt,smac} into a single and straightforward Application Programming Interface (API) \cite{api}. Thereby it creates a framework already integrating the infrastructure needed for processing data from different modalities with various data processing and learning algorithms, different strategies for hyperparameter optimization and state-of-the-art performance evaluation using nested cross validation.

In the following, we will first establish objectives for an Educational Machine Learning Code Generator and then describe a proof-of-concept software implementation; the PHOTON Wizard. Finally, the notion of Educational Machine Learning Code Generators and the PHOTON Wizard implementation are discussed in the context of applied-science ML and future developments and conclusions are outlined.  

\section{Methods}
\label{methods}
\subsection{Objectives for Educational Machine Learning Code Generators}

Code generators are tools that translate input information into valid source code \cite{codegeneration}. In the use case of educational Machine Learning infrastructure, we intend to utilize this process in order to lecture users on code implementation. In particular, GUI control operations shall be translated to appropriate source code capable of executing the desired action. While usability of GUIs in general and GUI control elements in particular is a widely researched field, we would like to add some objectives in regard to the educational component of such a code generating GUI system. 
\begin{description}

\item{\textbf{Structure.}} An object structure is needed which divides the problem into sub-tasks that can be presented as coherent GUI control elements and code units. In addition, the object structure educates the user on available choices, key components and best-practices.  

\item{\textbf{Feedback.}} The key to successful learning is the direct connection between GUI interactions and code output. User GUI interactions are represented in form of appropriate changes to the generated code.

\item{\textbf{Transparency.}} While implicit (hidden) default settings can foster readability and increase convenience in GUIs, an educational approach requires all information to be explicit, i.e. the source code must be transparently mapped to the GUI control elements. This is true especially for e.g. imports or other non-trivial settings across the entire target scope. 

\item{\textbf{Information.}} As the main goal of the system is to educate users, supplying explanations in the GUI is a key incentive. Specifically, the user should have direct access to background knowledge and further learning resources regarding the meaning of a particular concept, an algorithm or a parameter e.g. in form of explanative tooltips, educational texts, videos or links. 

\end{description}
\medskip

\subsection{The PHOTON Wizard}

\begin{figure*}[ht]
\vskip 0.2in
\begin{center}
\centerline{\includegraphics[width=\textwidth]{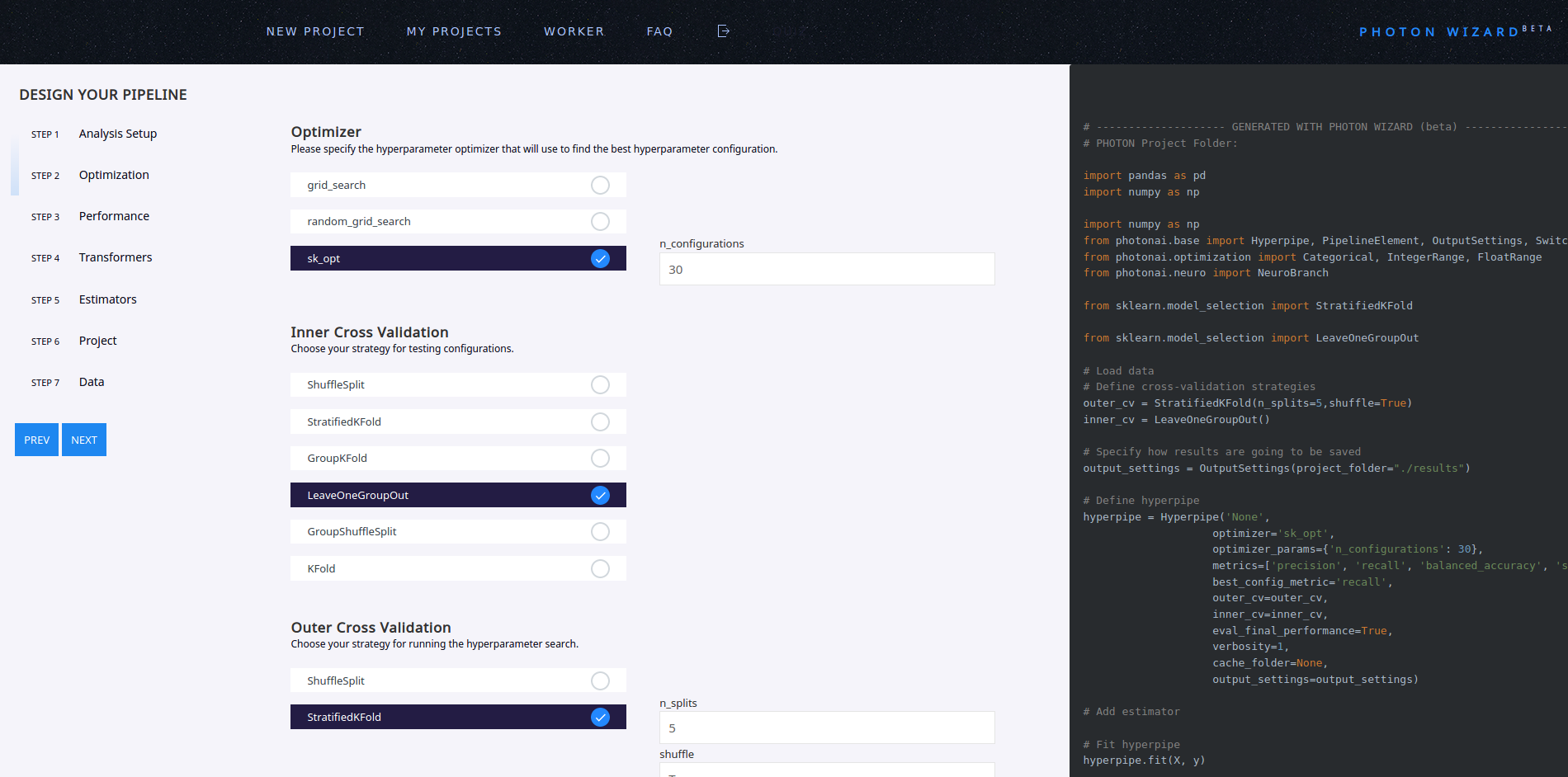}}
\caption{The PHOTON Wizard web application. Here, step two is depicted collecting information about  training, evaluation and optimization. The user chooses suitable hyperparameter optimization and cross validation strategies including their respective parameters. Simultaenously, Python source code is generated matched to the user's choices and GUI interactions. }
\label{fig:optimization}
\end{center}
\vskip -0.2in
\end{figure*}

In the following, we will introduce the PHOTON Wizard, a proof-of-concept implementation showcasing the idea of Educational Machine Learning Code Generators.

The PHOTON Wizard is a web application that enables the design of custom ML pipelines based on the PHOTON framework, a Python library functioning as high-level API to rapid ML model development \cite{photon}. PHOTON automatizes the repetitive development, training, hyperparameter optimization, and evaluation workflow and integrates several state-of-the-art toolboxes \cite{scikitlearn, scikitlearnapi,tensorflow,keras,imblearn,skopt,smac} covering data processing and learning algorithms, modality-specific functions, hyperparameter optimization strategies and other useful concepts such as data augmentation and class balancing strategies. Due to the wide scope of PHOTON, it is perfectly suitable to educate users on the development of machine learning models. 

\begin{figure}[ht]
\vskip 0.2in
\begin{center}
\centerline{\includegraphics[width=12cm]{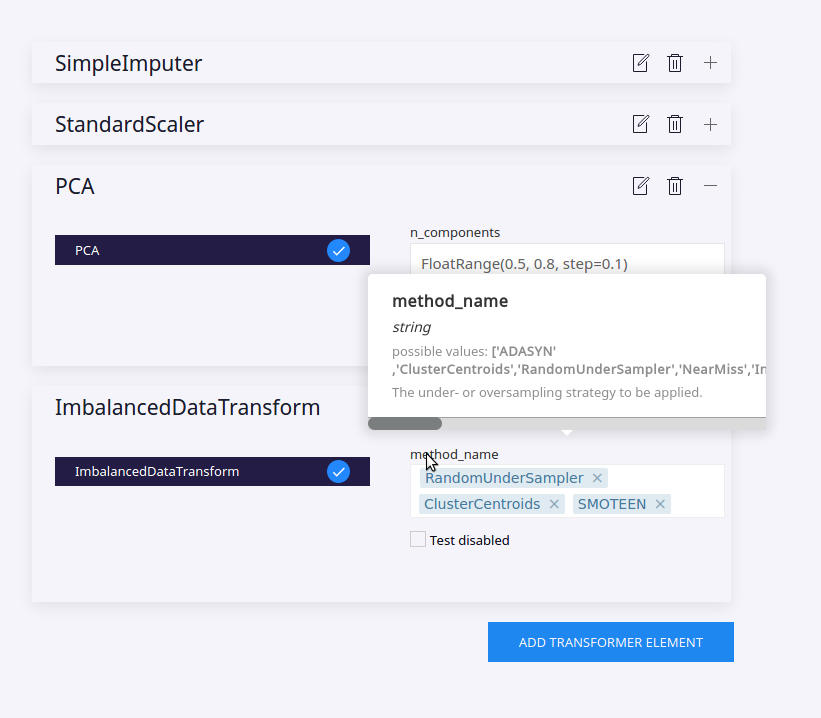}}
\caption{The PHOTON Wizard allows to dynamically design a machine learning pipeline. Here, graphical control elements (SelectBoxes and TextBoxes) are used to select and customize a sequence of data processing steps. The order of the items can be changed via drag-and-drop. When hovering over a particular data processing algorithm or a respective parameter, tooltips showcase further information.}
\label{transformers}
\end{center}
\vskip -0.2in
\end{figure}

In the PHOTON Wizard, we conceptualize an ML project as a basic scaffold that can be customized with building blocks, such as data processing and learning algorithms, hyperparameter optimization strategies, performance metrics etc., selectable from a variety of options. For example, hyperparameter optimization strategies are represented as control elements (SelectBoxes) and can be chosen from a predefined selection (see figure \ref{fig:optimization}). In addition, data processing steps can be selected, customized, ordered (see figure \ref{transformers}) and finally, one or more learning algorithms can be added to the pipeline. For each building block, default settings or hyperparameters are provided, which are internally adjusted to previous user choices. For example, performance metrics are filtered according to the prior specification of the analysis type (regression or classification) and the number of cross-validation folds is adjusted in relation to the number of training data available.

In addition, we divide the design process in steps that subsequently request user input in form of ML design choices. In accordance with the common terminology for stepwise assisted setup approaches, we name the application PHOTON Wizard \cite{wizard}. Implicitly, this type of user interface conveys information about how to break down complex tasks - here the development of a ML model - into a series of simpler steps.

Importantly, feedback is provided in form of changes in the generated source code reflecting the choice of GUI elements. In this way, the operational concepts of the GUI are directly translated into information on how to achieve the same action via PHOTON code. Each time the user finishes specifications in one of the steps, the source code adapts accordingly.  

Finally, a complete and fully functional code script is generated, including Python import statements as well as Python statements to load training and test data using the Python library pandas \cite{pandas}.  

In order to conveniently present educational information, each selectable item is associated with a tooltip containing a short explanation, helpful tips and links to further online resources. Also, aim and scope as well as important key considerations for each step of the ML workflow are shortly described at the top of the page.

The PHOTON Wizard is implemented in Flask \cite{flask}, a Python micro framework for web development adhering to the MVC Pattern \cite{mvc}. Data is stored in the open source document-oriented MongoDB database \cite{mongodb} and serialized using the pymodm library \cite{pymodm}. In order to facilitate the information exchange between GUI elements and database objects, we introduce a custom Binding layer \cite{webapps} that extracts the user input from the submitted requests and stores them in the corresponding data entities. In order to enhance usability for the selection and ordering of the data processing algorithms within the pipeline, we use a javascript plugin (JQuery Sortable \cite{jquery}) to dynamically drag and drop items.    

The content of the PHOTON Wizard, i.e. all available building blocks such as data processing and learning algorithms, hyperparameter optimization and cross validation strategies as well as performance metrics are easily adaptable via Excel files. Within these excel files, default parameters can be defined that nudge users to sensible default values  according to prior selections. We use a tag system to identify both suitable elements and default parameters, e.g. to limit the available performance metrics to those suitable for a regression task. Additional elements can be integrated via the PHOTON registry.

A demo of the PHOTON Wizard can be seen \href{https://wizard.photon-ai.com}{online}, while the source code is publicly available under the GNU General Public License v3.0 \cite{gplv3} on github. In order to facilitate hosting and deployment, we also include a docker image specification file.

\section{Discussion}
\label{discussion}

With the PHOTON Wizard, we introduce a GUI that generates valid Python source code to design, optimize and evaluate custom Machine Learning pipelines. We thereby enable non-programmers to apply Machine Learning and  - most importantly - by showing how GUI actions translate to source code foster the capability of implementing custom ML solutions. 

While other excellent educational resources exist, our approach more directly engages the users, helping to overcome typical limitations of a GUI and translate their often exceptional understanding of a problem domain and its technical concepts into practice. 

In the future, we would like to extend the GUI to include additional ML elements; particularly those specific to data modalities.

While providing users with an easily accessible tool to generate valid source code, an advanced learning task with e.g. a GPU-based, distributed computation approach, still requires in-depth knowledge about computational infrastructure, potentially excluding certain groups. To simplify this, future additions may include dedicated docker containers with pre-installed libraries etc.

\section{Conclusion}
\label{conclusion}
With the concept of Educational Machine Learning Code Generators, we aim to mitigate the trade-off between accessibility and flexibility in the context of ML software. Transforming GUI-based choices into valid, executable code in real-time combines easy accessibility with insights into the implementation of custom code. In addition, further content-related extensions are facilitated as they now only require changes to pre-generated code. Finally, seeing one’s choices in the GUI manifest into valid code in real-time has, in our experience, greatly increased user’s motivation to learn how to code.  We introduce the PHOTON Wizard as a concrete, proof-of-concept web application for building custom machine learning pipelines on top of the Python PHOTON framework. In addition to a theoretical didactic concept, we thus provide a hands-on implementation for anyone to try. With this, we hope to foster code understanding, convey basic knowledge about the machine learning workflow, integrate current best-practice for valid performance estimations and, exemplified in PHOTON, contribute to an intuitive understanding of how to develop custom solutions.


\bibliographystyle{unsrt}  
\bibliography{Wizard} 

\end{document}